\documentclass[letterpaper]{article} 
\usepackage{aaai2026}  
\usepackage{times}  
\usepackage{helvet}  
\usepackage{courier}  
\usepackage[hyphens]{url}  
\usepackage{graphicx} 
\usepackage{natbib}  
\usepackage{caption} 
\usepackage{algorithm}
\usepackage{algorithmic}

\usepackage{newfloat}
\usepackage{listings}
\definecolor{codegreen}{rgb}{0,0.6,0}
\definecolor{codegray}{rgb}{0.5,0.5,0.5}
\definecolor{codepurple}{rgb}{0.58,0,0.82}
\definecolor{backcolour}{rgb}{0.95,0.95,0.92}
\DeclareCaptionStyle{ruled}{labelfont=normalfont,labelsep=colon,strut=off} 
\floatstyle{ruled}
\newfloat{listing}{tb}{lst}{}
\floatname{listing}{Listing}

\usepackage[inline]{enumitem}
\usepackage{subcaption}
\usepackage{tikz}
\usetikzlibrary{decorations.pathreplacing,calligraphy,patterns}
\usepackage{tikz}
\usetikzlibrary{positioning}
\usetikzlibrary{backgrounds}
\usetikzlibrary{fit}
\usetikzlibrary{automata}
\usetikzlibrary{shapes.geometric}
\usetikzlibrary{decorations.pathreplacing,calligraphy}

\usepackage{amsmath}
\usepackage{amsthm}
\usepackage{thm-restate}
\usepackage{amssymb}
\usepackage{thmtools}
\usepackage{mathtools}
\usepackage[capitalise]{cleveref}
\usepackage[all]{foreign}
\usepackage{forest}
\usepackage{multicol}
\usepackage{lipsum}
\usepackage[flushleft]{threeparttable}
\usepackage{multirow}

\usepackage{packages/commands} 
\usepackage{packages/logic}    
\usepackage{packages/symbols}  

\usepackage{tabularx}
\usepackage{colortbl}

\newtheorem{theorem}{Theorem}
\newtheorem{definition}{Definition}
\newtheorem{proposition}{Proposition}

\newtheorem{mycondition}{Condition}
\crefname{mycondition}{Condition}{Conditions}

\newenvironment{claim}[1]{\par\noindent\underline{Claim:}\space#1}{}
\newenvironment{claimproof}[1]{\par\noindent\underline{Proof:}\space#1}{\hfill $\blacksquare$}

\newif\ifarxiv
\arxivtrue

\ifarxiv
\title{Automata-less Monitoring via Trace-Checking (Extended Version)}
\else
\title{Automata-less Monitoring via Trace-Checking}
\fi
\author{
  Andrea Brunello,
  Luca Geatti,
  Angelo Montanari,
  Nicola Saccomanno \\
}
\affiliations{
  Department of Mathematics, Computer Science and Physics\\
  University of Udine\\
  Via delle Scienze 206, Udine, Italy\\
  \{andrea.brunello, luca.geatti, angelo.montanari, nicola.saccomanno\}@uniud.it
}

\begin{document}

\maketitle

\begin{abstract}
  In runtime verification, \emph{monitoring} consists of  analyzing the
  current execution of a system and determining, on the basis of the 
  observed finite trace, whether all its possible continuations satisfy 
  or violate a given specification. 
  This is typically done by synthesizing a monitor--often
  a Deterministic Finite State Automaton (\DFA)--from logical specifications
  expressed in Linear Temporal Logic (\LTL) or in its finite-word variant
  (\LTLf).  Unfortunately, the size of the resulting \DFA may incur
  a doubly exponential blow-up in the size of the formula.

  In this paper, we identify some conditions under which monitoring can be
  done \emph{without} constructing such a \DFA. We build on the
  notion of \emph{intentionally safe and cosafe} formulas, introduced
  in [Kupferman \& Vardi, FMSD, 2001], to show that monitoring of these
  formulas can  be carried out through \emph{trace-checking}, that is, by
  directly evaluating them on the current system trace, with a 
  polynomial complexity in the size of both the trace and the formula.

  In addition, we investigate the complexity of recognizing intentionally safe
  and cosafe formulas for the safety and cosafety fragments of \LTL and
  \LTLf. As for \LTLf, we show that all  formulas in these fragments are
  intentionally safe and cosafe, thus removing the need for the check. 
  As for \LTL, we prove that the problem is in \PSPACE, significantly
  improving over the \EXPSPACE complexity of full \LTL.
\end{abstract}

\ifarxiv
\begin{links}
    \link{Conference version}{}
\end{links}
\else
\begin{links}
    \link{Extended version}{METTERE LINK AD ARXIV}
\end{links}
\fi

\section{Introduction}
\label{sec:intro}

Runtime verification~\cite{leucker2009brief} refers to a family of
techniques to check whether the execution of a system under scrutiny 
complies with the desired properties. Unlike model checking, which
exhaustively explores all possible system behaviors, runtime verification
analyzes only the current execution trace. Nevertheless, runtime
verification techniques offer substantial advantages: 
\begin{enumerate*}[label=(\roman*)]
  \item they do not require a complete model of the system, 
  \item they can be directly applied to the system implementation, and
  \item they are generally more efficient.
\end{enumerate*}

One of the most commonly used approaches in runtime verification is
\emph{monitoring}~\cite{leucker2009brief,bauer2011runtime}, which involves
the automatic construction of a \emph{monitor} from a logical property.
This monitor takes as input the current execution trace and outputs: 
\begin{enumerate*}[label=(\roman*)]
  \item $\top$ if all extensions of the trace satisfy the property;
  \item $\bot$ if all extensions violate the property; or 
  \item $?$ if the verdict is still undetermined. 
\end{enumerate*}
The monitor’s verdict is, by definition, irrevocable.  It follows that not
all logical properties are monitorable.  For example, it is not possible to
construct a monitor meeting the specified criteria for properties such as
``infinitely many times $p$.''

Typically, properties suitable for monitoring fall into two classes: 
\begin{enumerate*}[label=(\roman*)]
  \item
    \emph{safety properties}, which assert that ``something bad never
    happens'', and
  \item \emph{cosafety properties}, which assert that ``something good eventually
happens''. 
\end{enumerate*}
Every violation of a safety property contains a \emph{bad prefix}, \ie
a finite word such that all its extensions violate the property.
Conversely, every execution that satisfies a cosafety property contains
a \emph{good prefix}, \ie a finite word such that all its extensions
satisfy the property~\cite{kupferman2001model}.

The logical framework we consider to specify monitorable properties is
\LTL~\cite{pnueli1977temporal} and its finite-word counterpart
\LTLf~\cite{DeGiacomoV13}. Both extend propositional logic with temporal
modalities. As an example, the property $\ltl{G(\lnot deadlock)}$ is
a typical safety property, which states that the system never enters
a deadlock. An example of a cosafety property is $\ltl{(rules) U (goal)}$,
stating that the system eventually reaches a goal and respects certain
rules up to that point.

Two syntactic fragments of \LTL and \LTLf are particularly significant in
the context of safety and cosafety properties. The first one is \FpLTL,
which consists of formulas of the form $\ltl{F(\psi)}$, where $\ltl{F}$ is
the ``eventually'' modality and $\psi$ is a pure past formula, \ie one with
past temporal modalities only. Dually, the fragment \GpLTL consists of
formulas of the form $\ltl{G(\psi)}$, where $\ltl{G}$ is the ``always''
modality and $\psi$ is a pure past formula. It has been shown that a safety
(resp., cosafety) property is expressible in \LTL if and only if it is
expressible in \GpLTL (resp., in \FpLTL)~\cite{ChangMP92}.

A significant drawback of \LTL and \LTLf monitoring is that it typically
requires constructing a Deterministic Finite Automaton (\DFA) from the
input formula~\cite{leucker2009brief,bauer2011runtime}. Even for safety and
cosafety properties, this automaton can be doubly exponential in the size
of the formula~\cite{kupferman2001model}. This severely limits the
practical applicability of monitoring as the size of the formula increases.

In this paper, we investigate under which conditions monitoring for \LTL
and \LTLf can be reduced to \emph{trace-checking}--that is, deciding the
truth of a formula directly on the execution trace, without the need to
build a \DFA. To this end, we exploit the notions of \emph{intentionally
safe} and \emph{intentionally cosafe} formulas, introduced by Kupferman
and Vardi in the early 2000s in the context of model
checking~\cite{kupferman2001model}. Both notions are based on the concept
of \emph{informative models}. A model is informative for a formula $\phi$ if
it contains sufficient information to determine whether $\phi$ is true or
false. As an example, the word $\seq{\set{p}}$ is an informative model for the \LTL
formula $\ltl{F(p)}$, since $p$ holds at the first position. In contrast,
the same word is \emph{not} informative for $\ltl{F(p && (X q || X ! q))}$,
because evaluating the formula requires a position satisfying $p$ followed
by a position where either $q$ or $\lnot q$ holds, but $\seq{\set{p}}$ has
no successor. Notably, since $\ltl{X q || X ! q}$ is a tautology\footnote{%
  In~%
  \ifarxiv
    \cref{app:taut}%
  \else
    \cite[Section C]{brunello2025automata}%
  \fi%
  , we explain why the problem of determining whether
  a formula is intentionally (co)safe is not equivalent to removing
  tautological subformulas from the original formula, as the example above
  might misleadingly suggest.
}, the formulas $\ltl{F(p)}$ and $\ltl{F(p && (X q || X ! q))}$ are
logically equivalent, but only the former has $\seq{\set{p}}$ as an
informative model.

According to Kupferman and Vardi~\cite{kupferman2001model}, an \LTL formula
is intentionally safe (resp., intentionally cosafe) if it specifies a safety (resp., cosafety) property and all of its bad (resp., good)
prefixes are informative for the formula. As an example, $\ltl{F(p)}$ is
intentionally cosafe, because every good prefix includes a position where
$p$ holds. Conversely, $\ltl{F(p && (X q || X ! q))}$ is \emph{not}
intentionally cosafe, because $\seq{\set{p}}$ is a good prefix (all
extensions satisfy the formula), but it is not informative.

This paper makes three main contributions:

First, we prove that for all intentionally
cosafe (resp., safe) \LTL or \LTLf formulas, monitoring can be performed in an
\emph{automata-less} fashion. Specifically, the monitoring task can be
reduced to checking whether the formula is satisfied (resp., violated) by
the current trace (\emph{trace-checking} problem), which has two key
advantages: 
\begin{enumerate*}[label=(\roman*)]
  \item it does not require automaton construction, and 
  \item it can be solved in time $\O(|\sigma| \cdot |\phi|)$, where
    $\sigma$ is the trace and $\phi$ is the formula.
\end{enumerate*}

Second, we show that in the case of the \GpLTL and \FpLTL fragments of
\LTLf (\ie interpreted over finite words), every formula is intentionally
safe and intentionally cosafe, respectively. This result enables
monitoring of formulas in these fragments via trace-checking without
explicitly verifying the intentionally (co)safety condition.

Third, we present an algorithm to decide whether a formula in \GpLTL
(resp., \FpLTL), interpreted over infinite words, is intentionally safe
(resp., intentionally cosafe).  This serves as a prerequisite for enabling
the monitoring of such formulas without the construction of an
automaton--that is, through trace-checking, as described
in~\cref{sec:mon}. Our main technical result is that this problem
is in \PSPACE, as the algorithm operates in nondeterministic polynomial
space. This improves upon the known complexity of deciding intentionally
(co)safety for full \LTL, which is \EXPSPACE~\cite{kupferman2001model}.

The paper is structured as follows. \Cref{sec:back} provides background
knowledge. \Cref{sec:mon} shows the reduction of monitoring to
trace-checking for intentionally (co)safe formulas, and it proves that all
formulas in \GpLTL and \FpLTL, interpreted over finite words, are
intentionally safe and intentionally cosafe, respectively.
\Cref{sec:intentionally:infinite} presents a \PSPACE algorithm to check
whether \GpLTL and \FpLTL formulas over infinite words are intentionally
(co)safe. Finally, \cref{sec:conc} summarizes the achieved results and
outlines directions for future work.  The proofs of all the theorems are
reported in~%
\ifarxiv
  \cref{app:proofs}.
\else
  \cite[Section E]{brunello2025automata}.
\fi

\section{Background}
\label{sec:back}

In this section, we provide the necessary background.

Let $\Sigma=\set{a,b,c,\dots}$ be an alphabet, \ie a finite set of symbols.
A \emph{finite word} over $\Sigma$ is defined as a finite (possibly empty)
sequence of symbols in $\Sigma$. The empty word is denoted by $\epsilon$.
An \emph{infinite word} over $\Sigma$ is an infinite sequence of symbols in
$\Sigma$.  The set of all finite words over $\Sigma$ (including the empty one) 
is denoted by $\Sigma^*$, while the set of all infinite words over
$\Sigma$ is denoted by $\Sigma^\omega$. Additionally, we define $\Sigma^+
\coloneqq \Sigma^* \setminus \set{\epsilon}$.
A \emph{language of finite words} is a subset $\lang \subseteq \Sigma^*$.
Similarly, a \emph{language of infinite words} is a subset $\lang \subseteq
\Sigma^\omega$. The complement of a language $\lang$ is denoted by
$\overline{\lang}$.
Henceforth, the term \emph{property} will be used as a synonym for
language.
The \emph{length} of a word $\sigma$ is defined as follows: $|\sigma| = 0$
if $\sigma = \epsilon$; $|\sigma| = n$ if $\sigma = \langle \sigma_{0},
\ldots, \sigma_{n-1} \rangle \in \Sigma^{*}$; and $| \sigma | = \omega$ if
$\sigma \in \Sigma^{\omega}$.
Given $\sigma \in \Sigma^+ \cup \Sigma^\omega$, and a position $i
\in \set{0,\dots,|\sigma|-1}$, we define $\sigma_{[0,i]}$ as the prefix of
$\sigma$ up to position $i$.
We denote with $\sigma \cdot \sigma'$ the concatenation of the finite
word $\sigma$ with the finite or infinite word $\sigma'$.

\subsection{Linear Temporal Logic}
\label{sub:ltl}

We begin by introducing the syntax of \emph{Linear Temporal Logic} (\LTL)
\cite{pnueli1977temporal} and \emph{Linear Temporal Logic over finite
words} (\LTLf) \cite{DeGiacomoV13}. The two logics share the grammar.  From
now on, let $\AP = \set{p,q,r,\dots}$ be a set of atomic propositions.

\begin{definition}
\label{def:syntax:ltlf}
  A formula $\phi$ of \LTL or \LTLf over \AP is inductively defined as
  follows:
  \begin{align*}
    \phi \coloneqq 
      &p                     \choice
      \lnot p               \choice
      \phi_1 \lor \phi_2    \choice
      \phi_1 \land \phi_2   \choice \\
      &\ltl{X \phi}          \choice
      \ltl{wX \phi}         \choice
      \ltl{\phi_1 U \phi_2} \choice
      \ltl{\phi_1 R \phi_2} \choice \\
      &\ltl{Y \phi}          \choice
      \ltl{wY \phi}         \choice
      \ltl{\phi_1 S \phi_2} \choice
      \ltl{\phi_1 T \phi_2}
  \end{align*}
  %
\end{definition}

Temporal modalities $\ltl{X}$, $\ltl{wX}$, $\ltl{U}$, and $\ltl{R}$ are
called \emph{future} modalities.  Similarly, temporal modalities $\ltl{Y}$,
$\ltl{wY}$, $\ltl{S}$, and $\ltl{T}$ are called \emph{past} modalities. The
\emph{size} of $\phi$, denoted by $|\phi|$, is defined as the number of its
symbols.

Let us now define the semantics of \LTL and \LTLf. \LTL formulas are
interpreted over infinite words over the alphabet $2^\AP$, \ie over words
$\sigma \in (2^\AP)^\omega$, while  \LTLf formulas are interpreted over
finite and nonempty words over the alphabet $2^\AP$, \ie over words $\sigma
\in (2^\AP)^+$.

The satisfaction of a formula $\phi$ of \LTL (resp., \LTLf) by an
infinite word (resp., a finite, nonempty word) $\sigma
= \seq{\sigma_0,\sigma_1,\dots}$ at position $i$, denoted by $\sigma,i
\models \phi$, is inductively defined as follows:
\begin{itemize}
  \item $\sigma,i \models p$ (resp., $\lnot p$) iff $p\in \sigma_i$ (resp.,
    $p\not\in \sigma_i$);
  \item $\sigma,i \models \ltl{\phi_1 || \phi_2}$ iff $\sigma,i\models
    \phi_1$ or $\sigma,i \models \phi_2$;
  \item $\sigma,i \models \ltl{\phi_1 & \phi_2}$ iff $\sigma,i\models
    \phi_1$ and $\sigma,i \models \phi_2$;
  \item $\sigma,i \models \ltl{X \phi}$ iff $i+1<|\sigma|$ and $\sigma,i+1
    \models \phi$;
  \item $\sigma,i \models \ltl{wX \phi}$ iff $i+1=|\sigma|$ or $\sigma,i+1
    \models \phi$;
  \item $\sigma,i \models \ltl{\phi_1 U \phi_2}$ iff $\exists i\le
    j<|\sigma| \suchdot (\sigma,j \models \phi_2 \land \forall i\le
    k < j \suchdot \sigma,k \models \phi_1)$;
  \item $\sigma,i \models \ltl{\phi_1 R \phi_2}$ iff ($\forall j \ge
    i \suchdot \sigma, j \models \phi_2 ) \lor (\exists k \ge i \suchdot
    \sigma, k \models \phi_1 \land \forall i \le j \le k \suchdot \sigma,
    j \models \phi_2$);
  \item $\sigma,i \models \ltl{Y \phi}$ iff $i>0$ and $\sigma,i-1
    \models \phi$;
  \item $\sigma,i \models \ltl{wY \phi}$ iff $i=0$ or $\sigma,i-1
    \models \phi$;
  \item $\sigma,i \models \ltl{\phi_1 S \phi_2}$ iff $\exists j\le
    i \suchdot (\sigma,j \models \phi_2 \land \forall j<
    k \le i \suchdot \sigma,k \models \phi_1)$;
  \item $\sigma,i \models \ltl{\phi_1 T \phi_2}$ iff $(\forall 0 \le j \le
    i \suchdot \sigma, j \models \phi_2) \lor (\exists k \le i \suchdot
    \sigma, k \models \phi_1 \land \forall i \ge j \ge k \suchdot \sigma,
    j \models \phi_2)$.
\end{itemize}
We write $\sigma \models \phi$ to denote $\sigma,0\models\phi$. In such a
case, we say that $\sigma$ is a \emph{model} of $\phi$. The language of an
\LTL formula $\phi$ over \AP, denoted by $\lang(\phi)$, is the
set $\set{\sigma \in (2^\AP)^\omega \suchthat \sigma \models \phi}$.
Similarly, the language of an \LTLf formula $\phi$ over \AP, denoted by
$\langfin(\phi)$, is defined as the set $\set{\sigma \in (2^\AP)^+
\suchthat \sigma \models \phi}$.
Two \LTL (resp., \LTLf) formulas $\phi$, $\phi'$ are \emph{equivalent} if and only if
$\lang(\phi) = \lang(\phi')$ (resp., $\langfin(\phi) = \langfin(\phi')$).

\subsubsection{Fragments of \LTL and \LTLf.}
A number of derived modalities can be defined by using the basic ones of
\LTL and \LTLf. Among them, we would like to remind the following ones:
\begin{enumerate*}[label=(\roman*)]
  \item $\ltl{\true := p || ! p}$;
  \item $\ltl{F \phi := \true U \phi}$;
  \item $\ltl{G \phi := \false R \phi}$;
  \item $\ltl{O \phi := \true S \phi}$;
  \item $\ltl{H \phi := \false T \phi}$.
\end{enumerate*}
Temporal modalities $\ltl{F}$, $\ltl{G}$, $\ltl{O}$, and $\ltl{H}$ are
commonly referred to as \emph{eventually}, \emph{globally}, \emph{once},
and \emph{historically}, respectively.
Here is an intuitive account of their semantics:
\begin{itemize}
  \item $\ltl{F(p)}$ (resp., $\ltl{O(p)}$) reads \emph{``there exists
    a time point in the future (resp., past) where $p$ holds''};
  \item $\ltl{G(p)}$ (resp., $\ltl{H(p)}$) forces each position in the
    future (resp., past) to contain $p$.
\end{itemize}

In the following, we consider three specific fragments of \LTL and \LTLf,
which are characterized by the use of pure past temporal formulas. We
define the \emph{pure past fragment} of \LTL (and \LTLf), denoted by \pLTL,
as the set of \LTL (and \LTLf) formulas that do not contain any future
temporal modality. 

Pure past formulas $\phi$ of \pLTL are interpreted at the \emph{last} time
position of a finite, nonempty word, and we write $\sigma \models \phi$ if and only if
$\sigma,n \models \phi$ where $n = |\sigma|-1$. The language of a \pLTL
formula $\phi$ is defined as $\langfin(\phi) = \set{\sigma \in (2^\AP)^+
\suchthat \sigma \models \phi}$. As an example, the \pLTL formula $\ltl{Y
\true}$ is satisfied by all the finite, nonempty words of length at least
two.

On the basis of the pure past fragment of \LTL, as well as of \LTLf, the
logics \FpLTL and \GpLTL~\cite{ChangMP92,DBLP:conf/time/ArtaleGGMM23} are
defined as follows.

\begin{definition}
\label{def:Fpltl:Gpltl}
  \FpLTL is the set of formulas of the form $\ltl{F}(\psi)$ where $\psi$ is
  a \pLTL formula.
  \GpLTL is the set of formulas of the form $\ltl{G}(\psi)$ where $\psi$ is
  a \pLTL formula.
\end{definition}

As an example, the formula $\ltl{F(p & Y H q)}$ asks for the existence of
a time point in the future where $p$ holds, and forces $q$ to be true in
the prefix of the word up to that point. 

\subsection{Safety and cosafety properties}

Safety properties state the absence of undesirable behaviors,
typically phrased as \emph{ ``something bad never happens''}, e.g.,
the absence of deadlocks. Dually, cosafety properties guarantee
that \emph{``something good will eventually happen''}, such as the
reachability of a target state in a planning problem.


The definition of safety languages of infinite or finite, nonempty words is as follows.

\begin{definition}[Safety properties]
\label{def:safety}
  A language $\lang \subseteq \Sigma^\omega$ (resp., $\lang \subseteq
  \Sigma^+$) is \emph{safety} if, for all $\sigma \not\in \lang$, there
  exists a position $i \ge 0$ (resp., $i \in \set{0,\dots,|\sigma|-1}$)
  such that $\sigma_{[0,i]}\cdot\sigma'\not\in\lang$, for all $\sigma' \in
  \Sigma^\omega$ (resp., $\sigma' \in \Sigma^*$). The prefix
  $\sigma_{[0,i]}$ is called a \emph{bad prefix} for $\lang$.
\end{definition}

Cosafety languages are defined dually:  every word in the language must have a \emph{good prefix}, \ie a prefix such that all its
possible continuations belong to the language.

\begin{definition}[Cosafety properties]
\label{def:cosafety}
  A language $\lang \subseteq \Sigma^\omega$ (resp., $\lang \subseteq
  \Sigma^+$) is \emph{cosafety} if, for all $\sigma \in \lang$, there
  exists a position $i \ge 0$ (resp., $i \in \set{0,\dots,|\sigma|-1}$)
  such that $\sigma_{[0,i]}\cdot\sigma'\in\lang$, for all $\sigma' \in
  \Sigma^\omega$ (resp., $\sigma' \in \Sigma^*$). The prefix
  $\sigma_{[0,i]}$ is called a \emph{good prefix} for $\lang$.
\end{definition}

Given a formula $\phi$ of \LTL over the set of atomic propositions \AP, we
say that $\phi$ is a \emph{safe} (resp., \emph{cosafe}) formula if and only if
$\lang(\phi)$ is a safety (resp., cosafety) language over the alphabet
$2^{\AP}$.  The same applies to the case of \LTLf, considering
$\langfin(\phi)$ in place of $\lang(\phi)$.

It is immediate to observe that a language $\lang$ is safety if and
only if its complement $\overline{\lang}$ is cosafety. Owing to this
duality, cosafety properties are often preferred in practical applications,
as they are typically easier to handle.

In~\cite{ChangMP92}, it is proved that \GpLTL (resp., \FpLTL), when
interpreted on infinite words, captures exactly the set of all safety
languages (resp., all cosafety languages) that are definable in \LTL. The
same holds under finite-word
interpretation~\cite{DBLP:journals/lmcs/CimattiGGMT23}.

\begin{proposition}
\label{prop:synt:char:finite:safe}
  Let $\lang \subseteq \Sigma^\omega$ be a language definable in \LTL. It
  holds that $\lang$ is safety (resp., cosafety) if and only if $\lang
  = \lang(\phi)$, for some formula $\phi \in \GpLTL$ (resp., for some
  formula $\phi \in \FpLTL$). The same holds under finite-word
  interpretation.
\end{proposition}

\subsection{Automata on finite words}

We recall the standard definition of a Deterministic Finite Automaton
(\DFA~\cite{hopcroft2001introduction}).

\begin{definition}
\label{def:Dfa}
  A \emph{Deterministic Finite Automaton} (\DFA) is a tuple $\autom
  = \tuple{Q,\Sigma,q_0,\delta,F}$, where
  \begin{enumerate*}[label=(\roman*)]
    \item $Q$ is a finite set of states;
    \item $\Sigma$ is a finite alphabet;
    \item $q_0 \in Q$ is the initial state;
    \item $\delta : Q \times \Sigma \to Q$ is the transition function; and
    \item $F \subseteq Q$ is the set of final states.
  \end{enumerate*}
\end{definition}
Given a \DFA $\autom = \tuple{Q,\Sigma,q_0,\delta,F}$, we say that a word
$\sigma\in\Sigma^*$ is \emph{accepted} by $\autom$ if \emph{the} state
reached by $\autom$ reading $\sigma$ is final.  The language of $\autom$,
denoted by $\lang(\autom)$, is the set of words accepted by $\autom$. We
say that two \DFAs $\autom$ and $\autom'$ are equivalent if $\lang(\autom)
= \lang(\autom')$.
The size of $\autom$, written $|\autom|$, is defined as the cardinality of
$Q$.

All formulas in \LTLf and \pLTL can be effectively translated into a \DFA
that recognizes the same language. However, there is a fundamental
difference between the two settings: while for \LTLf the size of the
minimal equivalent \DFA is subject to a doubly-exponential lower bound in
the size of the formula~\cite{DeGiacomoV13}, for \pLTL there exist
algorithms that allow one to build an equivalent \DFA whose size is only exponential
in the size of the formula~\cite{de2021pure}.

\begin{proposition}
\label{prop:ltlf:dfa}
  For every formula $\phi$ of \LTLf, there exists a \DFA $\autom_\phi$ such
  that $\langfin(\phi) = \lang(\autom_\phi)$ and $|\autom_\phi| \in
  2^{2^{poly(|\phi|)}}$.
  For every formula $\phi$ of \pLTL, there exists a \DFA $\autom_\phi$ such
  that $\langfin(\phi) = \lang(\autom_\phi)$ and $|\autom_\phi| \in
  2^{poly(|\phi|)}$.
\end{proposition}

\subsection{Monitoring}
\label{sub:monitoring}
Monitoring is a lightweight runtime verification
technique~\cite{leucker2009brief}, which involves generating a monitor to
analyze an execution of the system under consideration either online (\ie
during runtime) or offline (\eg by processing the system's logs). The
monitor outputs either an inconclusive result (denoted by \textit{?}) or
a definitive verdict: a violation (resp., satisfaction) if all possible
continuations of the observed execution are bad (resp., good).

The \emph{monitor for a language of infinite words
$\lang\subseteq\Sigma^\omega$} (resp., \emph{for a language of finite words
$\lang\subseteq\Sigma^+$}) is defined as a function $\mon_{\lang}
: \Sigma^* \to \set{\top,\bot,\textit{?}}$ such that, for all $\sigma \in
\Sigma^*$,
\begin{align*}
  \mon_{\lang}(\sigma) \coloneqq
  \begin{cases}
    \top \mbox{ iff } \forall \sigma' \in \Sigma^\omega \text{ (resp., $\in
    \Sigma^*$)} : \sigma\cdot\sigma' \in \lang \\
    \bot \mbox{ iff } \forall \sigma' \in \Sigma^\omega \text{ (resp., $\in
    \Sigma^*$)} : \sigma\cdot\sigma' \not\in \lang \\
    \textit{?} \mbox{ otherwise}
  \end{cases}
\end{align*}
When the monitor returns $\top$, it indicates that all possible
continuations satisfy the property $\lang$; this corresponds to the case of
cosafety languages. Conversely, if the monitor returns $\bot$, it means
that an irremediable violation of $\lang$ has occurred, as in the case of
safety properties. Given an \LTL or \LTLf formula $\phi$, we will denote by
$\mon_\phi$ the monitor $\mon_{\lang(\phi)}$.
The \emph{monitoring problem} for \LTL and \LTLf is the problem of building a monitor
$\mon_{\phi}$ for a given \LTL or \LTLf formula $\phi$.

%

Given a temporal formula $\phi$, the classical approach to monitoring
involves building one \DFA for the good prefixes of $\phi$ and one for its
bad prefixes. The monitor basically consists of these two automata: as it
reads the input word, \ie the system execution, it produces the output
$\top$, if the automaton for the good prefixes has reached one of its final
states, or $\bot$, if the automaton for the bad prefixes has reached one of
its final states. In all other cases, the output is \textit{?}.

\section{Automata-less monitoring}
\label{sec:mon}

In this section, we identify the conditions under which monitoring of \LTL
and \LTLf formulas can be reduced to the \emph{trace-checking
problem}--that is, determining whether $\sigma$ is a model of $\phi$, where
$\sigma$ represents the system trace and $\phi$ is a formula. This
reduction eliminates the need to construct a deterministic automaton for
$\phi$, a process that incurs doubly-exponential complexity in the worst
case (\cref{prop:ltlf:dfa}).

To establish these conditions, we utilize the classification of safety
properties proposed in~\cite{kupferman2001model}, based on the notion of
\emph{intentionally safe formulas}, and extend this notion to
\emph{intentionally cosafe formulas}.

\subsection{Intentionally (co)safe formulas}

Intuitively, a finite word $\sigma \in (2^\AP)^+$ is an informative model
for a formula $\phi$ over $\AP$ if it contains all the information
necessary to determine the satisfaction of $\phi$ by $\sigma$.  For
instance, consider the \LTL formula $\phi \coloneqq \ltl{p & X p}$ and the
finite word $\sigma \coloneqq \seq{\set{p}}$. Although $\sigma$ satisfies
$p$ at the first (and only) position, it is not sufficient to conclude the
satisfaction of $\phi$ over an infinite extension of $\sigma$.  Indeed,
$\sigma$ can be extended by a trace beginning with a position in which
$p$ is false, yielding an infinite word that violates $\phi$. The same
considerations hold also for formulas of type $\ltl{G}(\psi)$ and for the
case of \LTLf formulas.
Below, we provide the definition of an informative model. 

\begin{definition}[Informative model in \LTL and \LTLf]
\label{def:informative:word:inf}
  Let $\sigma \coloneqq \seq{\sigma_0,\dots,\sigma_{n-1}} \in (2^\AP)^+$ be
  a finite word over $2^\AP$ and let $\phi$ be a formula of \LTL and \LTLf
  over \AP.  For all $0\le i<n$, we define $\sigma,i \models_B \phi$ as
  follows:
  \begin{enumerate}
    \item $\sigma,i \models_B \ltl{wX \psi}$ iff $\sigma,i \models_B
      \ltl{X \psi}$;
    \item $\sigma,i \models_B \ltl{\psi_1 R \psi_2}$ iff $\sigma,i
      \models_B \ltl{\psi_2 U (\psi_1 & \psi_2)}$;
    \item all the other cases are defined as in the definition of
      $\models$, up to replacing $\models$ with $\models_B$.
  \end{enumerate}
  We write $\sigma \models_B \phi$ iff $\sigma,0 \models_B \phi$.  We say
  that \emph{$\sigma$ is an informative model for $\phi$} iff $\sigma
  \models_B \phi$.
\end{definition}

In~
\ifarxiv
  \cref{app:kupf:vardi}%
\else
  \cite[Section D]{brunello2025automata}%
\fi%
, we discuss about the differences
between~\cref{def:informative:word:inf} and the original definition of
informative model given in~\cite{kupferman2001model}, and we show that they
are equivalent, up to considering $\phi$ in place of $\lnot\phi$ and the
use of $\models_B$ (which simplifies the proofs of the results) in place of
the labeling function $L$.

\paragraph{Examples.}
Consider the word $\sigma \coloneqq \seq{\set{p}}$. It holds that $\sigma$
is an informative model for the formula $\ltl{F(p)}$ because in the first
position $p$ is true. However, $\sigma$ is \emph{not} an informative model
for the formula $\ltl{F(p && (X q || X ! q))}$ because, at the first (and
only) position of $\sigma$, proposition $p$ holds but both $\ltl{X q}$ and
$\ltl{X !q}$ are false due to the absence of a successor position, \ie
$\sigma,0 \not\models \ltl{X q || X ! q}$ and thus $\sigma \not \models_B
\ltl{F(p && (X q || X ! q))}$. Intuitively, this means that $\sigma$ does
not provide sufficient information to determine the satisfaction of the
formula over any of its possible extensions.

As another example, consider the trace $\sigma \coloneqq
\seq{\set{p},\set{p},\set{p}}$ and the formula $\phi \coloneqq \ltl{F
G(p)}$, which is shorthand for $\ltl{\true U (\false R p)}$. It holds that,
for every position $i$ in $\sigma$, we have $\sigma, i \not\models_B
\ltl{\false R p}$, since, according to the definition of $\models_B$, this
would be equivalent to requiring that $\sigma, i \models \ltl{p U (p
& \false)}$. It follows that $\sigma$ is \emph{not} an informative model
for $\phi$. Intuitively, the reason is that the formula $\ltl{F G(p)}$
always requires an infinite trace to be satisfied, and no finite trace
(such as $\sigma$) can contain sufficient information to determine whether
it can be satisfied.

We define the \emph{trace-checking problem} for \LTL and \LTLf as follows.

\begin{definition}[Trace-checking problem]
\label{def:trace:check}
  Let $\sigma \in (2^\AP)^+$ be a word and let $\phi$ be an \LTL (or \LTLf)
  formula over the set of atomic propositions $\AP$.  The
  \emph{trace-checking problem} is the problem of establishing whether
  $\sigma \models_B \phi$.
\end{definition}

The time complexity of the trace-checking problem is polynomial in the size of
both the word and the formula. Specifically, given a word $\sigma \in
\Sigma^*$ and a formula $\phi$, the problem of deciding whether $\sigma
\models_B \phi$ holds can be solved in $\O(|\sigma| \cdot |\phi|)$
time~\cite{kupferman2001model,sistla1985complexity}.

Starting from the notion of informative models, \citet{kupferman2001model}
define the concept of \emph{intentionally safe formulas}, which we extend
here below to \emph{intentionally cosafe formulas}.

\begin{definition}[Intentionally safe formulas~\cite{kupferman2001model}]
\label{def:intentional:safe}
  A safe formula $\phi$ of \LTL (resp., \LTLf) is \emph{intentionally safe} iff
  all bad prefixes of $\lang(\phi)$ (resp., $\langfin(\phi)$) are also
  informative models for $\lnot \phi$.  
\end{definition}

\begin{definition}[Intentionally cosafe formulas]
\label{def:intentional:cosafe}
  A cosafe formula $\phi$ of \LTL (resp., \LTLf) is \emph{intentionally cosafe}
  iff all good prefixes of $\lang(\phi)$ (resp., $\langfin(\phi)$) are also
  informative models for $\phi$.  
\end{definition}

\paragraph{Examples.}
The \LTL formula $\phi \coloneqq \ltl{F(p && (X q || X ! q))}$ is
\emph{not} intentionally cosafe, because the finite word $\seq{\set{p}}$
is a good prefix (in fact, for every infinite continuation, the resulting
word satisfies the formula), but it is not an informative model.
The same holds for the \LTLf formula $\ltl{F(p && (F q || G ! q))}$.
Moreover, by a simple duality, it is easy to check that the \LTL formula
$\ltl{G(p || (X q && X ! q))}$ is \emph{not} intentionally safe.

On the contrary, formula $\phi' \coloneqq \ltl{F(p)}$ is intentionally
cosafe. Interestingly, $\phi$ and $\phi'$ are equivalent (they recognize
the same language).  As we will point out, this means that we \emph{cannot}
reduce monitoring of $\ltl{F(p && (X q || X ! q))}$ to trace-checking,
because, with the input trace $\sigma \coloneqq \seq{\set{p}}$, the
monitoring should output $\top$ whereas the trace-checking would return the
inconclusive verdict (\textit{?}).

In~\cite{kupferman2001model}, the notion of informative model is employed
to introduce a three-level classification of safe formulas, one of which
consists of the \emph{intentionally safe} formulas. Intuitively, this term
reflects the idea that the user has accurately expressed the intended
property, without introducing errors or redundant subformulas. 
For example, the formula $\ltl{G(p || (X q && X ! q))}$ is not
intentionally safe, as it contains the subformula $\ltl{X q && X ! q}$,
which evidently reflects a user error.

In this paper, we adopt the class of intentionally (co)safe formulas
from this classification as a sufficient condition for enabling
\emph{automata-less monitoring}--that is, monitoring that does not
require the construction of a \DFA and reduces directly to the
trace-checking problem (\cref{def:trace:check}).

\subsection{Automata-less monitoring via trace-checking}
The following theorem shows that, for intentionally (co)safe formulas, the
monitoring problem can be reduced to trace-checking, which admits
a polynomial-time solution. Intuitively, this holds because, for such
formulas, every good prefix encodes all the information necessary to
determine the truth of the formula over the system trace (and \viceversa).

\begin{restatable}{theorem}{montc}
\label{th:mon2tc}
  For all intentionally cosafe (resp., intentionally safe) formulas
  $\phi$ of \LTL and \LTLf over \AP, and for all finite words $\sigma \in
  (2^\AP)^+$, it holds that:
  \begin{align*}
    \mon_\phi(\sigma) = \top \ (\text{resp., } \bot)
      \quad \Iff \quad 
    \sigma \models_B \phi \ (\text{resp., } \lnot \phi)
  \end{align*}
\end{restatable}

\begin{figure}[t]
\centering
\begin{minipage}{0.25\textwidth}
  \begin{lstlisting}
fun mon_cosafe($\sigma$,$\phi$):
  if($\sigma \models_B \phi$):
    return $\top$;
  return $\textit{?}$;\end{lstlisting}
\end{minipage}%
\begin{minipage}{0.25\textwidth}
  \begin{lstlisting}
fun mon_safe($\sigma$,$\phi$):
  if($\sigma \models_B \lnot\phi$):
    return $\bot$;
  return $\textit{?}$;\end{lstlisting}
\end{minipage}
  \caption{Monitoring algorithm for intentionally (co)safe formulas based
  on trace-checking.}
\label{fig:mon2tc}
\end{figure}

Based on~\cref{th:mon2tc}, we present in~\cref{fig:mon2tc} the
trace-checking-based monitoring algorithm for intentionally (co)safe
formulas.
For intentionally cosafe formulas $\phi$, the algorithm returns $\top$ if
the input trace $\sigma$ is such that $\sigma \models_B \phi$ (\ie a good
prefix has been found); otherwise, it returns \textit{?}. For intentionally
safe formulas $\phi$, the algorithm returns $\bot$ if $\sigma \models_B
\lnot \phi$ (\ie a bad prefix has been found); otherwise, it returns
\textit{?}.

A key observation is that the satisfaction relation $\sigma \models_B \phi$
(as well as $\sigma \models_B \lnot\phi$) can be decided \emph{without}
constructing any automaton. In particular, the trace-checking procedure can
be performed in time $\O(|\sigma| \cdot |\phi|)$. This stands in contrast
to the automata-theoretic approach, where the automaton corresponding to
$\phi$ may be of doubly-exponential size in the worst case relative to the
size of $\phi$.

\subsection{Intentionality in $\mathsf{F(pLTL)}$ and $\mathsf{G(pLTL)}$ on
finite words}
\label{sub:intentionally:finite}

An important question concerns the computational complexity of determining
whether a given formula of \LTL or \LTLf is intentionally (co)safe.
In~\cite{kupferman2001model}, it is established that, in the case of \LTL,
this decision problem is in \EXPSPACE. In the following, we demonstrate
that the situation is substantially more favorable for the fragments \FpLTL
and \GpLTL. Specifically: 
\begin{enumerate*}[label=(\roman*)]
  \item under the finite-word semantics, \emph{every} formula in these
    fragments is trivially intentionally (co)safe (\cref{th:finite:coinf}),
    thereby rendering the verification procedure unnecessary;
  \item under the infinite-word semantics, the problem can be decided
    within \PSPACE (\cref{sec:intentionally:infinite}).
\end{enumerate*}

\begin{restatable}{theorem}{finitecoinf}
\label{th:finite:coinf}
  All formulas of \FpLTL (resp., of \GpLTL) interpreted over finite words
  are intentionally cosafe (resp., intentionally safe).
\end{restatable}

Intuitively, \cref{th:finite:coinf} follows from the observation that, for
these fragments, the $\models_B$ semantics
(\cref{def:informative:word:inf}) coincides with the standard \LTLf
semantics.

\Cref{th:finite:coinf} establishes that, for formulas belonging to the
fragments \FpLTL and \GpLTL of \LTLf, the ``intentionally (co)safety''
condition is always satisfied. Consequently, by~\cref{th:mon2tc}, the
monitoring problem for such formulas can be soundly reduced to the
trace-checking problem, entirely avoiding the need for automaton
construction.

\section{Intentionality in $\mathsf{F(pLTL)}$ and $\mathsf{G(pLTL)}$ on
infinite words}
\label{sec:intentionally:infinite}

In this section, we present a \PSPACE algorithm for deciding
intentionally (co)safety for formulas in the fragments \FpLTL and \GpLTL
under infinite-word semantics. As previously mentioned, the
complexity is significantly lower than in the general case, where the best
known algorithm for determining intentionally (co)safety for \LTL formulas
operates in \EXPSPACE.

When the fragments \FpLTL and \GpLTL are interpreted over infinite words,
the statement of~\cref{th:finite:coinf} no longer holds. In particular,
there exist formulas in \FpLTL (resp., \GpLTL) that are not intentionally
cosafe (resp., not intentionally safe).
As a counterexample, consider the formula $\ltl{F(Y p)}$ over
$\AP \coloneqq \set{p}$, which expresses that \emph{``there exists a future
position whose predecessor satisfies p''}. The finite word $\seq{\set{p}}$
constitutes a good prefix for $\ltl{F(Y p)}$, since for every infinite
continuation $\sigma' \in (2^{\AP})^\omega$, the concatenated word
$\seq{\set{p}} \cdot \sigma'$ satisfies $\ltl{F(Y p)}$; indeed, the
subformula $\ltl{Y p}$ is fulfilled at the second position ($i = 1$).
However, $\seq{\set{p}}$ is \emph{not informative} for $\ltl{F(Y p)}$, as
its unary length forces the $\ltl{F}$ operator to select $i = 0$. At this
position, $\ltl{Y p}$ evaluates to false, since the predecessor position
does not exist in the word $\seq{\set{p}}$ and thus $\seq{\set{p}} \not
\models_B \ltl{F(Y p)}$.

The remainder of this section is devoted to the proof of the following
theorem.

\begin{theorem}
\label{th:pspace:inf}
  Checking whether a formula of \FpLTL (resp., \GpLTL), interpreted over
  infinite words, is intentionally cosafe (resp., intentionally safe)
  is in \PSPACE.
\end{theorem}

In the following, we present the algorithm for the fragment \FpLTL. The
case of \GpLTL can be handled analogously by exploiting the standard
duality argument between safety and cosafety properties.\footnotemark

\footnotetext{%
  It is straightforward to show that a formula of the form $\ltl{G(\psi)}$
  is intentionally safe if and only if its negation is intentionally
  cosafe. Consequently, to handle formulas of the form $\ltl{G(\psi)}$,
  it suffices to verify whether $\ltl{F(! \psi)}$ is intentionally
  cosafe.
}

The algorithm that we propose for deciding intentional cosafety of
a formula of the form $\ltl{F(\psi)}$ in \FpLTL proceeds in two steps:
\begin{itemize}
  \item \emph{Step 1}. Construct a \DFA $\autominf$ that recognizes the set of finite
    models that are informative for $\ltl{F(\psi)}$.  
  \item \emph{Step 2}. For all non-final states $q$ of $\autominf$,
    \emph{mark $q$ as final} iff there is a good prefix of $\ltl{F(\psi)}$
    inducing $\autominf$ to $q$. Intuitively, if this is the case, then
    there is at least one good prefix which is \emph{not} informative, so
    $\ltl{F(\psi)}$ is not intentionally cosafe.  Otherwise,
    $\ltl{F(\psi)}$ is intentionally cosafe.
\end{itemize}
Conceptually, as we will discuss, this procedure implements the
verification that the set of good prefixes of $\ltl{F(\psi)}$ is included
within the set of informative models of $\ltl{F(\psi)}$
(\cf~\cref{def:intentional:cosafe}).  To achieve \PSPACE complexity, we argue that
the second step must be carried out \emph{on-the-fly} during the
construction of the automaton in the first step.

\subsection{Step 1: Computing the informative models of $\mathsf{F(pLTL)}$}
The first step consists of constructing a \DFA, denoted $\autominf$, that
recognizes all informative models of $\phi$. This construction reduces to
building the \DFA for the \FpLTL formula interpreted over finite words,
since, as established by the following proposition, for \FpLTL formulas,
the notion of informative models (\cf~\cref{def:informative:word:inf})
coincides with the standard semantics of \FpLTL over finite words.

\begin{restatable}{proposition}{propinfmodelfinite}
\label{prop:inf:model:finite}
  For all formulas $\ltl{F(\psi)}$ in \FpLTL, and for all finite words
  $\sigma \in (2^\AP)^+$, it holds that $\sigma$ is an informative model for
  $\ltl{F(\psi)}$ iff $\sigma\models\ltl{F(\psi)}$ (under finite-word
  semantics).
\end{restatable}

Crucially, while for general formulas of \LTL over finite words the
automaton construction causes a doubly exponential blow-up
(\cref{prop:ltlf:dfa}), for the case of \FpLTL this can be avoided.
Intuitively, this is because, under finite-word semantics, $\ltl{F(\psi)}$
is equivalent to $\ltl{O(\psi)}$, which is a pure past formula;
thus,~\cref{prop:ltlf:dfa} can be leveraged to obtain an efficient
automaton construction.

\begin{restatable}{proposition}{propautomfpsi}
\label{prop:autom:fpsi}
  There exists a \DFA recognizing the language $\langfin(\ltl{F(\psi)})$ of
  size $2^{poly(n)}$, where $n=|\ltl{F(\psi)}|$.
\end{restatable}

We denote by $\autominf$ the \DFA constructed in~\cref{prop:autom:fpsi},
which, by~\cref{prop:inf:model:finite}, recognizes the set of informative
models of $\ltl{F(\psi)}$ and its size is at most exponential in
$|\ltl{F(\psi)}|$.

\begin{figure}[t]
    \centering
    \begin{subfigure}[b]{0.45\textwidth}
        \centering
        \begin{tikzpicture}[shorten >=1pt,node distance=2.5cm,on grid,auto,
            scale=0.8, every node/.style={scale=0.8}]
            \tikzstyle{every state}=[fill={rgb:black,1;white,10}]
            \node[state,initial,initial text=]    (s_0)     {$q_0$};
            \node[state]  (s_1)  [right of=s_0]   {$q_1$};
            \node[state,accepting]  (s_2)  [right of=s_1]   {$q_2$};

            \path[->]
            (s_0) edge [loop above] node[left,align=center, outer sep=6pt] {$\emptyset$} ()
            (s_0) edge[] node[align=center] {$\set{p}$} (s_1)
            (s_1) edge[] node[align=center] {$\emptyset,\set{p}$} (s_2)
            (s_2) edge [loop above] node[right,align=center, outer sep=6pt]
                {$\emptyset,\set{p}$} ();
        \end{tikzpicture}
        \caption{The \DFA $\mathcal{A}^{\mathit{inf}}_{\mathsf{F}(\mathsf{Y} p)}$.}
        \label{fig:phi:example:a}
    \end{subfigure}
    \hfill
    \begin{subfigure}[b]{0.45\textwidth}
        \centering
        \begin{tikzpicture}[shorten >=1pt,node distance=2.5cm,on grid,auto,
            scale=0.8, every node/.style={scale=0.8}]
            \tikzstyle{every state}=[fill={rgb:black,1;white,10}]
            \node[state,initial,initial text=] (q_0) {$q_0$};
            \node[state,accepting]  (q_1)  [right of=q_0]   {$q_1$};
            \node[state,accepting]  (q_2)  [right of=q_1]   {$q_2$};

            \path[->]
            (q_0) edge [loop above] node[left,align=center, outer sep=6pt] {$\emptyset$} ()
            (q_0) edge[] node[align=center] {$\set{p}$} (q_1)
            (q_1) edge[] node[align=center] {$\emptyset,\set{p}$} (q_2)
            (q_2) edge [loop above] node[right,align=center, outer sep=6pt]
                {$\emptyset,\set{p}$} ();
        \end{tikzpicture}
        \caption{The \DFA $\mathcal{A}^{\mathit{gp}}_{\mathsf{F}(\mathsf{Y} p)}$.}
        \label{fig:phi:example:b}
    \end{subfigure}
    \caption{The \DFAs (a) $\mathcal{A}_{\mathit{inf}}$ and (b)
      $\mathcal{A}_{\mathit{gp}}$ for the formula $\mathsf{F}(\mathsf{Y}p)$.}
    \label{fig:phi:example}
\end{figure}

\Cref{fig:phi:example:a} depicts the \DFA $\autominf$ for the
informative models of the formula $\ltl{F(Y p)}$. As previously noted, this
automaton does not recognize all good prefixes. In particular, it does not
accept the good prefix $\seq{\set{p}}$.

\subsection{Step 2: Marking the states in $\mathcal{A}_{\mathit{inf}}$}
The second step of the algorithm consists in what follows: starting from
the automaton $\autominf$, we examine each of its states
$q$ and mark $q$ as \emph{final} if and only if the following condition
holds.

\begin{mycondition}
\label{cond:gp}
  For every infinite word $\sigma \in (2^{AP})^\omega$, there exists an
  index $i \geq 0$ such that, starting from $q$, the automaton
  $\autominf$ reaches a final state after reading the first
  $i$ symbols of $\sigma$.
\end{mycondition}

We denote by $\automgp$ the \DFA obtained in this way, that is, after all
states have been examined. As an example, consider again the formula
$\ltl{F(Y p)}$. \cref{fig:phi:example:b} shows the \DFA $\automgp$ for the
formula $\ltl{F(Y p)}$, obtained from the \DFA
in~\cref{fig:phi:example}(a) by marking as final those states that
fulfill~\cref{cond:gp}--in this case, only the state $q_1$. It is easy to
see that, in this example, $\automgp$ accepts all and only the good
prefixes of $\ltl{F(Y p)}$, in particular the word $\seq{\set{p}}$, which
is not accepted by $\autominf$.

The following lemma proves that $\automgp$ accepts exactly the set of good
prefixes of $\ltl{F}(\psi)$.

\begin{restatable}{lemma}{automgplemma}
\label{lemma:automgp}
  The automaton $\automgp$ recognizes the set of good prefixes of the
  formula $\ltl{F(\psi)}$.
\end{restatable}

Crucially, since $\automgp$ has the same number of states and the same
number of transitions as $\autominf$, its size is the same as the size of
$\autominf$, that is, at most singly-exponential in the size of
$\ltl{F(\psi)}$ (\cref{prop:autom:fpsi}).

Therefore, to check whether $\ltl{F(\psi)}$ is intentionally cosafe, it
suffices to check whether the language of $\automgp$ is included in the
language of $\autominf$. The following proposition is a direct consequence
of~\cref{lemma:automgp,def:intentional:cosafe}.

\begin{proposition}
\label{prop:lang:equiv}
  The formula $\ltl{F(\psi)}$ is intentionally cosafe iff
  $\lang(\automgp) \subseteq \lang(\autominf)$.
\end{proposition}

\subsection{PSPACE Complexity of the Algorithm}
In this section, we demonstrate that verifying $\lang(\automgp)
\subseteq \lang(\autominf)$ as stated in~\cref{prop:lang:equiv} can be achieved
within the \PSPACE complexity class, by circumventing the explicit
construction of $\automgp$ in the first instance. Specifically, we will
show that it suffices to check whether a state of $\autominf$
satisfies~\cref{cond:gp} \emph{on-the-fly} with the construction of
$\autominf$.
Intuitively, this follows from the subsequent two observations:
\begin{enumerate}[label=(\roman*)]
  \item Since both automata are \emph{deterministic} and given that
    $\automgp$ is derived from $\autominf$ by eventually designating
    a subset of its states as final, marking a non-final state (which is
    also reachable from the initial state) as final implies that
    $\lang(\automgp) \not\subseteq \lang(\autominf)$.
  \item Verifying~\cref{cond:gp} for a state $q$ is equivalent to not
    finding a lasso-shaped path starting from $q$ and visiting only
    non-final states. Therefore, \cref{cond:gp} reduces to a reachability
    problem, which can be performed on-the-fly in \emph{nondeterministic
    logarithmic space} (\NL).
\end{enumerate}

\begin{figure}[t]
\centering
\begin{minipage}{0.5\textwidth}
\begin{lstlisting}
while( building the DFA $\autominf$ )
  for( q state of $\autominf$ )
    q' $\gets $ nondeterministically guess a state of $\autominf$
    if( q,q' are not final and q$\neq$q' and
        q is reachable from the initial state)
      if( q' is reachable from q and from 
          itself and all the states in 
          between are non-final )
             leave q as non-final
      else
             mark q as final
\end{lstlisting}%
\end{minipage}%
\caption{A \PSPACE algorithm for deciding whether $\mathsf{F}(\psi)$ is
  intentionally cosafe.}
\label{alg:gp}
\end{figure}

The pseudocode implementing the steps outlined above is presented
in~\cref{alg:gp}.  Intuitively, it checks whether~\cref{cond:gp} holds for
a given state $q$ of $\autominf$ by verifying the existence of
a lasso-shaped path starting from $q$ that visits only non-final states. If
such a path exists, then~\cref{cond:gp} is violated and $q$ is left as
non-final. Otherwise, $q$ is marked as a new final state.

The algorithm in~\cref{cond:gp} runs in polynomial space (\PSPACE).  While
constructing $\autominf$--which, according to~\cref{prop:autom:fpsi}, has
a size at most exponential in $|\ltl{F(\psi)}|$--we can simultaneously
execute a reachability check (in \NL) to determine if at least one state of
$\autominf$ satisfies~\cref{cond:gp}. If such a state exists, then
$\lang(\automgp) \not\subseteq \lang(\autominf)$, and
by~\cref{prop:lang:equiv}, the formula $\ltl{F(\psi)}$ is not intentionally
cosafe. Conversely, after examining all states of $\autominf$ without
finding any state satisfying~\cref{cond:gp}, we can conclude that
$\ltl{F(\psi)}$ is intentionally cosafe. This approach yields a \PSPACE
complexity for the algorithm.


\begin{restatable}{lemma}{automgpalgorithm}
  The algorithm in~\cref{alg:gp} decides whether $\ltl{F(\psi)}$ is
  intentionally cosafe in nondeterministic polynomial space.
\end{restatable}

Since by Savitch's Theorem~\cite{DBLP:journals/jcss/Savitch70}, $\NPSPACE
= \PSPACE$, we have that the problem of deciding whether a formula of
\FpLTL (resp., of \GpLTL) is intentionally cosafe (resp., intentionally
safe) is in \PSPACE, proving~\cref{th:pspace:inf}.  This is in sharp
contrast with the \EXPSPACE upper bound proven in~\cite{kupferman2001model}
for deciding intentionally (co)safe formulas in full \LTL.

\section{Conclusions and Future Work}
\label{sec:conc}

In this paper, we exploited the notions of intentionally (co)safe formulas,
introduced in~\cite{kupferman2001model}, to reduce the monitoring
problem to trace-checking,
which is solvable in polynomial time. We showed that all formulas in the
fragments \FpLTL and \GpLTL over finite words are intentionally cosafe and
intentionally safe, respectively, and we gave a \PSPACE algorithm to
check intentionality in these fragments over infinite words, improving the
best-known doubly exponential upper bound for full \LTL.

This work yields several implications for the field of runtime
verification. As a matter of fact, by~\cref{prop:synt:char:finite:safe},
any (co)safety property expressible in \LTL or \LTLf can be equivalently
rewritten in the form \GpLTL or \FpLTL, respectively. In the case of \LTL,
intentional (co)safety can be verified using~\cref{alg:gp}: if the formula
satisfies this condition, automata-less monitoring is applicable.  For
finite words, our result establishes that all formulas within the \FpLTL
and \GpLTL fragments are inherently intentionally (co)safe. Consequently,
monitoring can be performed directly through trace-checking, in an
automata-less fashion.
In fact, in~\cite{DBLP:journals/corr/abs-2508-17786}, we demonstrated that
replacing the monitoring algorithm with a trace-checking approach yields
a substantial performance improvement in a framework combining monitoring
and machine learning for early failure detection. In this framework, past
system executions are analyzed using a variation of \FpLTL/\GpLTL formulas
interpreted over finite words (specifically, their real-time extension,
\ie fragments of \STL).


As for future work, identifying syntactic fragments of \LTL where all
formulas are intentionally (co)safe remains a significant open problem.

\appendix
\section{Acknowledgments}
The authors acknowledge the support from the 2024 Italian INdAM-GNCS
project \lq\lq Certificazione, monitoraggio, ed interpretabilità in sistemi
di intelligenza artificiale\rq\rq, ref. no. CUP E53C23001670001.
Luca Geatti, Angelo Montanari and Nicola Saccomanno acknowledge the support
from the Interconnected Nord-Est Innovation Ecosystem (iNEST), which
received funding from the European Union Next-GenerationEU (PIANO NAZIONALE
DI RIPRESA E RESILIENZA (PNRR) – MISSIONE 4 COMPONENTE 2, INVESTIMENTO 1.5
– D.D. 1058 23/06/2022, ECS00000043).
In addition, Angelo Montanari acknowledges the support from the MUR PNRR
project FAIR - Future AI Research (PE00000013) also funded by the European
Union Next-GenerationEU.

\section{Ethical Statement}
This manuscript reflects only the authors’ views and opinions, neither the
European Union nor the European Commission can be considered responsible
for them.

\ifarxiv
\section{Recognizing intentionally (co)safety is different from removing
tautological subformulas}
\label{app:taut}

It is important to emphasize that establishing whether a formula is
intentionally (co)safe is not equivalent to removing tautological
subformulas, contrary to what the example formula $\ltl{F(p && (X q ||
X ! q))}$ may suggest. In fact, consider the following cosafe formula taken
from~\cite{kupferman2001model}:
\[ \ltl{\phi := F(q) && F(r) && F((q || G F(! p)) && (r || G F(p)))} \]
The finite word $\seq{\set{q},\set{r}}$ is a good prefix as every infinite
extension of it satisfies either $\ltl{G F(p)}$ (\emph{``there exists
infinitely many time points in which $p$ is true''}) or $\ltl{G F(! p)}$
(\emph{``there exists infinitely many time points in which $\lnot p$ is
true''}).  However, it is not informative for the formula, as no finite
word can confirm whether $p$ (or $\lnot p$) holds infinitely often.  Thus,
$\phi$ is not intentionally cosafe, despite containing no tautological
subformulas.

\section{Differences between our definition of informative models and the
original one}
\label{app:kupf:vardi}

In~\cite{kupferman2001model}, Kupferman and Vardi define
a prefix as informative for $\phi$ if it contains all the information needed
to establish a \emph{violation} of $\phi$. This choice stems from their exclusive
focus on \emph{safety properties}, where such a characterization naturally
applies. By contrast, our work deals with both safety and cosafety
properties, and for the latter one seeks good prefixes, that is, prefixes whose
extensions all satisfy $\phi$. Hence, we found it more natural to define
informativeness directly with respect to $\phi$.

Another key difference is that our definition avoids the use of the
labeling function $L$. $L$ allows one to encode a particular semantics of
\LTL (namely, a modification of the satisfaction relation $\models$), but
we found it more convenient--evidenced by the brevity of our main
proofs--to define informativeness using $\models_B$.

One final distinction is that, whereas Kupferman and Vardi provide the
definition of intentionally safe formulas solely for the case of future
temporal operators, we extend this definition to also include past
operators. Conceptually, this extension comes at no additional cost, since
the past is inherently bounded and cannot be extended indefinitely, unlike
the future.

Below, we show that the two definitions are, in fact, equivalent, up to
considering $\phi$ in place of (the \emph{Negation Normal Form} of)
$\lnot\phi$ and not considering past operators.

We first report the original definition of informative model proposed
in~\cite{kupferman2001model}. To distinguish this definition from the one
that we give in this paper, we refer to it as \emph{``KV-informative
model''}. Before giving this definition, we define the {Negation Normal
Form of $\lnot\phi$}, denoted as $\nnf(\lnot\phi)$, as the formula
equivalent to $\lnot\phi$ obtained by pushing the negations only in front
of propositional atoms. We define the \emph{closure of a formula $\phi$},
denoted as $\closure(\phi)$, as the set of all its subformulas $\psi$ and
their negation $\nnf(\lnot\psi)$.

\begin{definition}{KV-informative model~\cite{kupferman2001model}}
\label{def:kv:informative}
  Let $\sigma \coloneqq \seq{\sigma_0,\dots,\sigma_{n-1}} \in (2^\AP)^+$ be
  a finite word over $2^\AP$ and let $\phi$ be a formula of \LTL and \LTLf
  over \AP.  We say that \emph{$\sigma$ is a KV-informative model for
  $\phi$} iff there exists a labeling function $L : {0,\dots,n} \to
  2^{\closure(\nnf(\lnot\phi))}$ such that the following hold:
  \begin{itemize}
    \item $\nnf(\lnot \phi) \in L(0)$;
    \item $L(n) = \emptyset$;
    \item for all $0\le i<n$ and for all $\psi \in L(i)$, the following
      hold:
    \begin{itemize}
      \item if $\psi = p$ (with $p\in \AP$, then $p \in \sigma_i$;
      \item if $\psi = \lnot p$ (with $p \in \AP$, then $p \in \sigma_i$;
      \item if $\psi = \phi_1 \lor \phi_2$, then either $\psi_1 \in L(i)$
        or $\psi_2 \in L(i)$;
      \item if $\psi = \phi_1 \land \phi_2$, then $\psi_1 \in L(i)$ and
        $\psi_2 \in L(i)$;
      \item if $\psi = \ltl{X \psi_1}$, then $\psi_1 \in L(i+1)$;
      \item if $\psi = \ltl{wX \psi_1}$, then $\psi_1 \in L(i+1)$;
      \item if $\psi = \ltl{\psi_1 U \psi_2}$, then either $\psi_2 \in
        L(i)$ or ($\psi_1 \in L(i)$ and $\ltl{\psi_1 U \psi_2} \in L(i+1)$);
      \item if $\psi = \ltl{\psi_1 R \psi_2}$, then $\psi_2 \in L(i)$ and
        ($\psi_1 \in L(i)$ or $\ltl{\psi_1 R \psi_2} \in L(i+1)$).
    \end{itemize}
  \end{itemize}
\end{definition}

The following proposition proves the correspondence between our definition
of informative model (\cref{def:informative:word:inf}) and the original one
proposed in~\cite{kupferman2001model}.

\begin{proposition}
\label{prop:informative:equiv}
  For all $\sigma = \seq{\sigma_0,\dots,\sigma_{n-1}} \in (2^\AP)^+$ and
  for all pure future formulas $\phi$ of \LTL and \LTLf, it holds that
  $\sigma$ is a KV-informative model for $\phi$ iff $\sigma$ is informative
  for $\nnf(\lnot \phi)$.
\end{proposition}
\begin{proof}
  We prove that, for all $\psi \in \closure(\nnf(\lnot\phi))$ and for all
  $0\le i<n$:
  \begin{align*}
    (\psi \in L(i) \land L(n) = \emptyset) \Iff \sigma,i \models_B \psi
  \end{align*}
  We proceed by induction on the structure of $\psi$.

  If $\psi = p$ (with $p\in\AP$), then $p \in L(i)$ and $L(n) = \emptyset$
  is equivalent to $p \in \sigma_i$ and $L(n) = \emptyset$, which in turn
  is equivalent to $\sigma,i \models_B p$.
  
  The cases for $\psi = \lnot p$, $\psi = \psi_1 \lor \psi_2$, and $\psi
  = \psi_1 \land \psi_2$ are straightforward to prove.

  Let $\psi = \ltl{X \psi_1}$. If $\ltl{X \psi_1} \in L(i)$ and $L(n)
  = \emptyset$, this means that $i$ \emph{cannot} be equal to $n-1$ and
  that $\psi_1 \in L(i+1)$. By inductive hypothesis, we have that
  $\sigma,i+1 \models_B \psi_1$. Together with the fact that $i+1<n$, it
  follows that $\sigma,i \models_B \ltl{X \psi_1}$.  If $\sigma,i \models_B
  \ltl{X \psi_1}$, then by definition of $\models_B$ we have that $i+1<n$
  and $\sigma,i+1 \models_B \psi_1$. By inductive hypothesis, we have that
  $\psi_1 \in L(i+1)$ and $i+1<n$. It follows that $\ltl{X \psi_1} \in
  L(i)$ and $L(n) = \emptyset$.

  The case for $\psi = \ltl{wX \psi_1}$ is proved in the same way as the
  case for $\psi = \ltl{X \psi_1}$.

  Let $\psi = \ltl{\psi_1 U \psi_2}$. If $\ltl{\psi_1 U \psi_2} \in L(i)$
  and $L(n) = \emptyset$, then the label $\ltl{\psi_1 U \psi_2}$ cannot be
  postponed after position $n-1$ (because $L(n) = \emptyset$). It follows
  that there exists an index $j$ between $i$ (included) and $n$ (excluded)
  such that $\psi_2 \in L(j)$ and $\psi_1 \in L(k)$ for all $k$ between $i$
  (included) and $j$ (excluded). By inductive hypothesis:
  \begin{align*}
    \exists i\le j<n \suchdot (\sigma,j \models_B \psi_2 \land \forall i\le
    k<j \suchdot \sigma,j \models_B \psi_1)
  \end{align*}
  which, by definition of $\models_B$, amounts to $\sigma,i \models_B
  \ltl{\psi_1 U \psi_2}$.
  Now, for the opposite direction, suppose that $\sigma,i \models_B
  \ltl{\psi_1 U \psi_2}$. By definition of $\models_B$, this is equivalent
  to:
  \begin{align*}
    \exists i\le j<n \suchdot (\sigma,j \models_B \psi_2 \land \forall i\le
    k<j \suchdot \sigma,j \models_B \psi_1)
  \end{align*}
  By inductive hypothesis, we have that:
  \begin{align*}
    \exists i\le j<n \suchdot (\psi_2 \in L(j) \land \forall i\le k<j
    \suchdot \psi_1 \in L(k))
  \end{align*}
  This implies that $L(n) = \emptyset$ and $\ltl{\psi_1 U \psi_2} \in
  L(i)$.

  The case for $\psi = \ltl{\psi_1 R \psi_2}$ can be proved with the same
  principle as the previous case.

  Choosing $\psi = \lnot \phi$ and $i=0$, we have that $\sigma$ is
  a KV-informative model for $\phi$ iff $\sigma,0 \models_B
  \nnf(\lnot\phi)$ (\ie $\sigma$ is informative for $\nnf(\lnot\phi)$).
\end{proof}

\section{Proofs of theorems}
\label{app:proofs}

\montc*
\begin{proof}
  We prove the case of intentionally cosafe formulas of \LTL. The safety
  case directly follows by duality.

  Let $\phi$ be an intentionally cosafe formula of \LTL over $\AP$ and
  let $\sigma \in (2^\AP)^+$. We first prove that $\mon_\phi(\sigma)
  = \top$ implies $\sigma \models_B \phi$.  If $\mon_\phi(\sigma) = \top$
  then, by definition of monitor, $\sigma \cdot \sigma' \models \phi$, for
  all $\sigma' \in (2^\AP)^\omega$.  By~\cref{def:cosafety}, this means
  that $\sigma$ is a good prefix of $\lang(\phi)$. Since, by hypothesis,
  $\phi$ is intentionally cosafe, it holds that all good prefixes are
  informative models, and thus $\sigma \models_B \phi$.

  We now prove that $\sigma \models_B \phi$ implies $\mon_\phi(\sigma)
  = \top$. First, we prove by induction that, for all formulas $\psi$ of
  \LTL, for all $\sigma \in (2^\AP)^+$ and for all $i\ge 0$, if $\sigma,i
  \models_B \psi$ then $\sigma\cdot\sigma',i \models \psi$ for all $\sigma'
  \in (2^\AP)^\omega$.  We divide in cases:

  \underline{Case $\psi = p \in \AP$.} If $\sigma,i \models_B p$, then, by
  definition of $\models_B$, we have that $p \in \sigma_i$, which in turn
  implies that $\sigma \cdot \sigma',i \models p$, for all $\sigma' \in
  (2^\AP)^\omega$. The same proofs works for also for the case $\psi
  = \lnot p$.

  \underline{Case $\psi = \ltl{\psi_1 | \psi_2}$.} If $\sigma,i \models_B
  \psi_1 \lor \psi_2$, then, by definition of $\models_B$, we have that
  either $\sigma,i \models_B \psi_1$ or $\sigma,i \models_B \psi_2$. By
  inductive hypothesis, we have that either $\sigma\cdot\sigma',i \models
  \psi_1$ for all $\sigma'\in(2^\AP)^\omega$, or $\sigma\cdot\sigma'',i
  \models \psi_2$ for all $\sigma'' \in (2^\AP)^\omega$. This implies that
  either $\sigma\cdot\sigma',i \models \psi_1$ or
  $\sigma\cdot\sigma',i\models \psi_2$, for all $\sigma'\in(2^\AP)^\omega$,
  which is equivalent to $\sigma\cdot\sigma',i \models \psi_1 \lor \psi_2$,
  for all $\sigma' \in (2^\AP)^\omega$. The same proof works also for the
  case $\psi = \psi_1 \land \psi_2$.
  
  \underline{Case $\psi = \ltl{Y \psi_1}$.} If $\sigma,i \models_B \ltl{Y
  \psi_1}$, then, by definition of $\models_B$, we have that $i>0$ and
  $\sigma,i-1 \models_B \psi_1$. By inductive hypothesis, $i>0$ and
  $\sigma\cdot\sigma',i-1 \models \psi_1$, for all $\sigma' \in
  (2^\AP)^\omega$, which is equivalent to
  $\sigma\cdot\sigma',i\models\ltl{Y\psi_1}$, for all $\sigma' \in
  (2^\AP)^\omega$. The same proof works also for the case $\psi = \ltl{wY
  \psi_1}$.

  \underline{Case $\psi = \ltl{\psi_1 S \psi_2}$.} If $\sigma,i \models_B
  \ltl{\psi_1 S \psi_2}$, then, by definition of $\models_B$, we have that
  $\exists j \le i (\sigma,j \models_B \psi_2 \land \forall j<k\le
  i (\sigma,k \models_B \psi_1))$. By inductive hypothesis we have that
  $\exists j \le i \suchdot \forall
  \sigma'\in(2^\AP)^\omega(\sigma\cdot\sigma',j \models \psi_2 \land
  \forall j<k\le i \suchdot \forall
  \sigma''\in(2^\AP)^\omega(\sigma\cdot\sigma'',k \models \psi_1))$. In
  turn, this is equivalent to $\forall \sigma'\in(2^\AP)^\omega \suchdot
  \exists j \le i (\sigma\cdot\sigma',j \models \psi_2 \land \forall j<k\le
  i (\sigma\cdot\sigma',k \models \psi_1))$. This is equivalent to $\sigma
  \cdot \sigma',i \models \ltl{\psi_1 S \psi_2}$, for all $\sigma' \in
  (2^\AP)^\omega$. The same proof works also for the case $\psi
  = \ltl{\psi_1 T \psi_2}$.

  \underline{Case $\psi = \ltl{X \psi_1}$}. If $\sigma,i \models_B \ltl{X
  \psi_1}$, then by definition of $\models_B$ we have that $i+1 < |\sigma|$
  and $\sigma,i+1 \models_B \psi_1$. By inductive hypothesis, $i+1
  < |\sigma|$ and $\sigma\cdot\sigma',i+1 \models \psi_1$ for all $\sigma'
  \in (2^\AP)^\omega$, which is equivalent to $\sigma\cdot\sigma',i \models
  \ltl{X \psi_1}$, for all $\sigma' \in (2^\AP)^\omega$.

  \underline{Case $\psi = \ltl{wX \psi_1}$}. If $\sigma,i \models_B \ltl{wX
  \psi_1}$, then by definition of $\models_B$ we have that $\sigma,i
  \models_B \ltl{X \psi_1}$, which in turn means that $i+1<|\sigma|$ and
  $\sigma,i+1 \models_B \psi_1$. By inductive hypothesis, we have that
  $i+1<|\sigma|$ and $\sigma\cdot\sigma',i+1 \models_B \psi_1$, for all
  $\sigma' \in (2^\AP)^\omega$. This implies that $\sigma\cdot\sigma',i
  \models \ltl{wX \psi_1}$, for all $\sigma' \in (2^\AP)^\omega$.

  \underline{Case $\psi = \ltl{\psi_1 U \psi_2}$.} If $\sigma,i \models_B
  \ltl{\psi_1 U \psi_2}$, then by definition of $\models_B$ we have that
  \begin{align*}
    &\exists i\le j<|\sigma| \suchdot (\sigma,j \models_B \psi_2 \land \\
    &\qquad \forall i\le k < j \suchdot \sigma,k \models_B \psi_1).
  \end{align*}
  By inductive hypothesis, we have that:
  \begin{align*}
    &\exists i\le j<|\sigma| \suchdot \forall \sigma' \in (2^\AP)^\omega
    (\sigma \cdot \sigma',j \models \psi_2 \land \\
    &\qquad \forall i\le k < j \suchdot \forall \sigma'' \in (2^\AP)^\omega
    (\sigma\cdot\sigma'',k \models \psi_1))
  \end{align*}
  which is equivalent to:
  \begin{align*}
    &\forall \sigma' \in (2^\AP)^\omega \suchdot \exists i\le j<|\sigma|
    (\sigma \cdot \sigma',j \models \psi_2 \land \\
    &\qquad \forall i\le k < j (\sigma\cdot\sigma'',k \models \psi_1))
  \end{align*}
  that is, $\sigma\cdot\sigma',i \models \ltl{\psi_1 U \psi_2}$, for all
  $\sigma' \in (2^\AP)^\omega$.

  \underline{Case $\psi = \ltl{\psi_1 R \psi_2}$.} If $\sigma,i \models_B
  \ltl{\psi_1 R \psi_2}$, then by definition of $\models_B$ we have that
  $\sigma,i \models_B \ltl{\psi_2 U (\psi_1 & \psi_2)}$. By inductive
  hypothesis, this means that:
  \begin{align*}
    \exists j<|\sigma| &\suchdot \forall \sigma'\in(2^\AP)^\omega (\sigma
    \cdot \sigma',j \models (\psi_1 \land \psi_2) \\
    &\land \forall i \le
    k < j \suchdot \forall \sigma'' \in (2^\AP)^\omega (\sigma \cdot
    \sigma'',k \models \psi_2))
  \end{align*}
  which is equivalent to:
  \begin{align*}
    \forall \sigma'&\in(2^\AP)^\omega \suchdot \exists j<|\sigma| 
      (\sigma \cdot \sigma',j \models (\psi_1 \land \psi_2) \\ 
      &\land \forall i \le k < j \suchdot (\sigma \cdot \sigma'',k \models
      \psi_2))
  \end{align*}
  which implies that $\sigma\cdot\sigma',i \models \ltl{\psi_1 R \psi_2}$,
  for all $\sigma' \in (2^\AP)^\omega$.

  This concludes the proof by induction.

  Therefore, since $\sigma \models_B \phi$, we have that $\sigma \cdot
  \sigma' \models \phi$, for all $\sigma' \in (2^\AP)^\omega$.  By definition of monitor
  (\cref{sub:monitoring}), this means that $\mon_\phi(\sigma) = \top$.

  The proof for the case of intentionally safe formulas follows the same
  principle. The proof for the case of \LTLf formulas is obtained by the
  proof above by considering $\langfin(\phi)$ instead of $\lang(\phi)$ and
  $\sigma' \in (2^\AP)^*$ instead of $\sigma' \in (2^\AP)^\omega$.
\end{proof}

\finitecoinf*
\begin{proof}
  We begin with the case of \FpLTL.  Let $\ltl{F(\psi)}$ be a \FpLTL
  formula, over the set of atomic propositions \AP, interpreted over finite
  words.

  First, notice that the set of good prefixes of $\ltl{F(\psi)}$ is exactly
  the set of \emph{models} of $\ltl{F(\psi)}$.

  \begin{claim}
    The set of good prefixes of $\ltl{F(\psi)}$ is exactly the set of
    models of $\ltl{F(\psi)}$.
  \end{claim}
  \begin{claimproof}
    To prove that every good prefix of $\ltl{F(\psi)}$ is a model of
    $\ltl{F(\psi)}$ it suffices to notice that, by definition, a good
    prefix $\sigma$ is such that $\sigma\cdot\sigma'\models \ltl{F(\psi)}$,
    for all $\sigma' \in (2^\AP)^*$. For the particular case in which
    $\sigma' = \epsilon$ (\ie the empty word), it follows that $\sigma
    \models \psi$, and thus that $\sigma \models \ltl{F(\psi)}$.

    To prove the opposite direction, consider a model $\sigma$ of
    $\ltl{F(\psi)}$. By definition of the $\ltl{F}$ operator and since
    $\psi$ is a pure past formula, it holds that there exists a prefix
    $\sigma'$ of $\sigma$ such that $\sigma'\cdot \sigma'' \models \psi$,
    for all $\sigma'' \in (2^\AP)^*$. Since $\sigma'$ is a prefix of
    $\sigma$, this implies that $\sigma \cdot \sigma'' \models
    \ltl{F(\psi)}$, for all $\sigma'' \in (2^\AP)^*$. Therefore, $\sigma$
    is a good prefix of $\ltl{F(\psi)}$.
  \end{claimproof}

  Now, to prove that all good prefixes of $\ltl{F(\psi)}$ are also
  informative, it suffices to notice that,
  by~\cref{def:informative:word:inf}, the relation $\models_B$ coincides
  with the relation $\models$ for the $\ltl{F}$ operator and for all the
  past operators. Therefore, since by the claim above all good prefixes
  $\sigma$ of
  $\ltl{F(\psi)}$ are also models of the formula, that is, $\sigma \models
  \ltl{F(\psi)}$, we have also that $\sigma \models_B \ltl{F(\psi)}$, that
  is, $\sigma$ is informative for $\ltl{F(\psi)}$. This concludes the proof
  that $\ltl{F(\psi)}$ is intentionally cosafe.

  The case of \GpLTL is obtained analogously by exploiting the fact that
  $\ltl{! G(\psi) == F(!\psi)}$ and the fact that the bad prefixes of
  $\ltl{G(\psi)}$ are the good prefixes of $\ltl{F(!\psi)}$.
\end{proof}

\propinfmodelfinite*
\begin{proof}
  Suppose that $\sigma$ is an informative model for $\ltl{F(\psi)}$.
  By~\cref{def:informative:word:inf}, this is equivalent to $\sigma
  \models_B \ltl{F(\psi)}$. Since the formula $\ltl{F(\psi)}$ features only
  the $\ltl{F}$ operator and past temporal operators,
  by~\cref{def:informative:word:inf}, we have that $\sigma \models_B
  \ltl{F(\psi)}$ iff $\sigma \models \ltl{F(\psi)}$, that is, $\sigma$ is
  a model of $\ltl{F(\psi)}$ under finite-word semantics.
\end{proof}

\propautomfpsi*
\begin{proof}
  It is trivial to prove that, under finite-word semantics,
  $\langfin(\ltl{F(\psi)}) = \langfin(\ltl{O(\psi)})$.
  By~\cref{prop:ltlf:dfa}, it follows that there exists a \DFA recognizing
  the language $\langfin(\ltl{O(\psi)})$ (and thus also the language
  $\langfin(\ltl{F(\psi)})$) of size $2^{poly(n)}$, where $n
  = |\ltl{O(\psi)}|$. Since the size of $\ltl{O(\psi)}$ is also the size of
  $\ltl{F(\psi)}$, the claim follows.
\end{proof}

\automgplemma*
\begin{proof}
  We first prove that every word accepted by $\automgp$ is a good prefix of
  $\ltl{F(\psi)}$. Let $\sigma$ be a word accepted by $\automgp$. By
  definition of acceptance, reading $\sigma$, $\automgp$ reaches a final
  state $q_f$. We divide in cases depending on whether $q_f$ is also final
  in $\autominf$.
  
  \underline{Case 1.} Suppose that $q_f$ is final in $\autominf$. Since
  $\autominf$ accepts the language (of finite words) of $\ltl{F(\psi)}$, it
  means that $\sigma \models \ltl{F(\psi)}$.  By the semantics of the
  $\ltl{F}$ operator and since $\psi$ is a pure past formula, we have that
  there exists a $i$ (with $0\le i < |\sigma|$) such that $\sigma_{[0,i]}
  \models \psi$.  Clearly, since $\psi$ is pure past, we have
  $\sigma_{[0,i]} \cdot \sigma',i \models \psi$, for all $\sigma' \in
  (2^\AP)^\omega$. Since $\sigma_{[0,i]}$ is a prefix of $\sigma$, this
  implies that $\sigma\cdot\sigma',i \models\psi$, for all
  $\sigma'\in(2^\AP)^\omega$.  This is equivalent to say that
  $\sigma\cdot\sigma' \models \ltl{F(\psi)}$, for all
  $\sigma'\in(2^\AP)^\omega$, and thus $\sigma$ is a good prefix of
  $\ltl{F(\psi)}$.

  \underline{Case 2.} Suppose that $q_f$ is \emph{not} final in
  $\autominf$. By definition of $\automgp$ (\cref{cond:gp}),
  this means that:
  \begin{align*}
    &\forall \sigma' \in (2^\AP)^\omega \suchdot
    \exists i \geq 0 \suchdot \ 
    \sigma \cdot \sigma'_{[0,i]} \text{ forces } \automgp \\
    &\text{ to reach a final state of } \autominf
  \end{align*}
  Since $\autominf$ recognizes the language
  $\langfin(\ltl{F(\psi)})$, we have that:
  \begin{align*}
    \forall \sigma' \in (2^\AP)^\omega \suchdot
    \exists i \geq 0 \suchdot \ 
    \sigma \cdot \sigma'_{[0,i]} \models \ltl{F(\psi)}
  \end{align*}
  By the semantics of the $\ltl{F}$ operator over finite- and infinite-word
  semantics, this implies that:
  \begin{align*}
    &\forall \sigma' \in (2^\AP)^\omega \suchdot
    \exists i \geq 0 \suchdot 
    \forall \sigma'' \in (2^\AP)^\omega \suchdot \\
    &\sigma \cdot \sigma'_{[0,i]} \cdot \sigma'' \models \ltl{F(\psi)}
  \end{align*}
  which is equivalent to:
  \begin{align*}
    \forall \sigma' \in (2^\AP)^\omega \suchdot
    \sigma \cdot \sigma' \models \ltl{F(\psi)}
  \end{align*}
  meaning that $\sigma$ is a good prefix of $\ltl{F(\psi)}$.

  We now prove the opposite direction, \ie every good prefix of
  $\ltl{F(\psi)}$ is accepted by $\automgp$.  Let $\sigma$ be a good prefix
  of $\ltl{F(\psi)}$, that is, $\sigma\cdot\sigma' \models \ltl{F(\psi)}$,
  for all $\sigma'\in(2^\AP)^\omega$.  By definition of the $\ltl{F}$
  operator, this means that, for all $\sigma' \in (2^\AP)^\omega$, there
  exists a $i\ge 0$, such that, $\sigma \cdot \sigma',i \models
  \ltl{\psi}$.  Let $\sigma'\in(2^\AP)^\omega$, let $i \ge 0$, and let $q$
  be the state reached by $\automgp$ after $i$ steps reading
  $\sigma\cdot\sigma'$.  Suppose by contradiction that $q$ is not final in
  $\automgp$. By~\cref{cond:gp}, this means that there exists an infinite
  word $\sigma''\in(2^\AP)^\omega$ such that, for all $i\ge 0$, all the
  states reached by $\automgp$ reading $\sigma''$ starting from $q$ are
  non-final in $\autominf$. In turn, this means that, for all
  $i\ge 0$, $\sigma \cdot \sigma'',i \not\models \psi$, and thus $\sigma
  \cdot \sigma'' \not\models \ltl{F(\psi)}$, implying that $\sigma$ is not
  a good prefix of $\ltl{F(\psi)}$, which contradicts our hypothesis.
\end{proof}

\automgpalgorithm*
\begin{proof}
  We first prove that a state $q$ is marked as final by~\cref{alg:gp} iff,
  for all infinite words $\sigma \in (2^\AP)^\omega$, there exists a $i\ge
  0$ such that, starting from $q$, $\autominf$ reaches a final state after
  $i$ steps reading $\sigma$.

  Consider the following condition: for all $\sigma \in (2^\AP)^\omega$,
  there exists a $i\ge 0$ such that $\autominf$ reaches a final state after
  $i$ steps. It is equivalent to: it is not true that there exists
  a $\sigma \in (2^\AP)^\omega$ such that, for all $i\ge 0$, $\autominf$
  reaches a non-final state after $i$ steps. In turn, this is equivalent to
  say that, in $\autominf$, starting from $q$, there exists an infinite
  path made only of non-final states. Since the number of states in
  $\autominf$ is finite, this is true iff there exists a non-final node
  reachable from $q$ and reachable from itself and all the states inbetween
  are non-final, which is exactly the definition of Lines $4$-$5$
  in~\cref{alg:gp}. Therefore, a node is marked as final (Line $8$)
  iff~\cref{cond:gp} is fulfilled.

  Since Lines $4$-$5$ amounts to a reachability check, which can be done in
  nondeterministic logarithmic space, and since the construction of
  $\autominf$ requires $2^{poly(n)}$ space (with $n=|\ltl{F(\psi)}|$) and
  can be done on-the-fly with the reachability check, the total complexity
  is nonderministic polynomial space.
\end{proof}
\fi

\bibliography{biblio}

\end{document}